# Community detection based on significance optimization in complex networks

Ju Xiang[1], Zhi-Zhong Wang[2], Hui-Jia Li[3,4,*], Yan Zhang[5], Fang Li[1], Li-Ping Dong[1], Jian-Ming Li[1, *]

[1] Department of Anatomy, Histology and Embryology, Neuroscience Research Center, Changsha Medical University, Changsha, 410219, Hunan, China.

[2] South City College, Hunan First Normal University, Changsha, 410205, Hunan, China.

[3] School of Management Science and Engineering, Central University of Finance and Economics, Beijing 100080, China.

[4] Academy of Mathematics and Systems Science, Chinese Academy of Sciences, Beijing 100190, China.

[5] Department of Computer Science, Changsha Medical University, Changsha, 410219, Hunan, China.

* Corresponding authors: Hui-Jia Li or Yan Zhang or Jian-Ming Li.

E-mail: xiang.ju@foxmail.com (J.X.); xiangju@aliyun.com (J.X.); hjli@amss.ac.cn (H.J.L.); zhangyancsmu@foxmail.com (Y.Z.); ljming0901@sina.com (J.M.L.)

**Abstract:** Community structure is an important structural property that extensively exists in various complex networks. In the past decade, much attention has been paid to the design of community-detection methods, but analyzing the behaviors of the methods is also of interest in the theoretical research and real applications. Here, we focus on an important measure for community structure, *significance* [Sci. Rep. 3 (2013) 2930]. Specifically, we in detail study the effect of various network parameters on this measure, analyze the critical behaviors of it in partition transition, and analytically give the formula of the critical points and the phase diagrams. The results shows that the critical number of communities in partition transition increases dramatically with the difference between inter- and intra-community link densities, and thus significance optimization displays higher resolution in community detection than many other methods, but it is also easily to lead to the excessive splitting of communities. By Louvain algorithm for significance optimization, we confirmed the theoretical results on artificial and real-world networks, and give a series of comparisons with some classical methods.

**PACS:** 89.75.–k; 89.75.Fb; 89.75.Hc

**Keywords:** Complex networks; Community detection; Resolution; significance

## CONTENTS

## 1. Introduction

Complex networks provide a kind of effective approach for understanding the structure and function of various complex systems in real world, such as the metabolic networks and protein-protein interaction networks [1]. In the past decade, many common topological properties were discovered and investigated widely in the complex networks, such as clustering, degree correlation and community structure [1, 2], which implies the existence of possible organization principles in the systems. The appearance of community structure means that the complex networks generally consist of groups of vertices within dense inner connections and sparse external connections, called communities or modules [1]. Community structure in complex networks is closely related to real functional grouping in real-world systems [3-5] and it can affect such dynamic processes as information diffusions and synchronizations [6, 7]. For example, Yan et al recently found that local targeted immunization outperforms global targeted immunization, if there exists apparent community structure in a network [8]; Wu et al shown that the abundance of communities in the social network can evidently foster the formation of cooperation under strong selection [9]. Therefore, a large number of methods have been proposed to detect the communities in complex networks based on various approaches, such as spectral analysis [10, 11], random walk dynamics [12-14], phase dynamics[15, 16], diffusion dynamics[17], label propagation [18-20], statistical models [21, 22], structural perturbation [23] and modularity optimization [24-26] (see refs [1, 27, 28] for reviews).

Much attention has been paid to the design of community-detection methods, while there is only a few works in analyzing the behaviors of the methods. Studying the behaviors of the methods is also of interest in the theoretical research and real applications. On the one hand, it could be helpful for understanding the method themselves, because the methods also have the scope of application themselves, though they are helpful for detecting and analyzing the structures of complex networks. On the other hand, it could promote the improvement of the methods or the development of more effective methods. For example, methods based on modularity optimization and Bayesian inference were found that there exist phase transitions from detectable to undetectable structures in community detection, which provides a bound on the achievable performance of the methods [29-31]. Botta et al presented a detailed analysis of modularity density, showing its superiors and drawbacks [32]. The limits of modularity, such as the resolution limit [33-35], implied the possible existence of multi-scale structures in networks, and promoted the proposal of various (improved) methods, especially the multi-resolution modularity or Hamiltonians [36-40]. Various approaches have been used to improve the performance of modularity-based methods [41-43]. Lai et al proposed the improved modularity-based method by random walk network preprocessing [42], and then enhanced the modularity-based belief propagation method by using the correlation between communities to improve the estimate of number of communities [43]. Chakraborty et al proposed a new post-processing technique by which many existing community-detection methods for hard partitions can be extended to soft partitions, based on the resemblance between identified non-overlapping and actual overlapping community structure [44].

Optimizing quality functions for community structures is a kind of popular strategy for community detection, such as *modularity* [24-26, 45-47], *Hamiltonians* [21], and "fitness" functions [48, 49]. In Ref [50], Traag et al proposed an important measure for community detection, called *significance*. It can be used to estimate the quality of community structures, by looking at how likely dense communities appear in random networks, and is defined as,

$$S = \sum_s \binom{n_s}{2} D(p_s \parallel p)$$
$$= \sum_s \frac{n_s(n_s-1)}{2}\left( p_s \ln\frac{p_s}{p} + (1-p_s)\ln\frac{1-p_s}{1-p} \right),$$
(1)

Here the sum runs over all communities; the density of community $s$, $p_s$, is the ratio of the number of existing edges to the maximum in the community; the density of network, $p$, is the ratio of the number of existing edges to the maximum in the whole network. This measure was initially proposed to determine significant scale of community structures, while it cloud also be directly optimized as objective function to find the optimal community partitions. And as reported, it shown excellent performance in may tests [50].

In this paper, we will analyze the effect of various network parameters on this measure, study in detail the critical behaviors of it in partition transition, and analytically give the formula of the critical points and the phase diagrams. By the Louvain algorithm, we confirmed the theoretical results on artificial and real-world networks, and give a series of comparisons with classical methods, including *Infomap, Walktrap, OSLOM, LP* and *modularity*. Finally, we come to conclusion.

## 2. Method

In this section, we firstly introduce a set of model networks (see Figure 1) and the analytic expression of *significance* in the networks, then analyze the relation between *significance* and various network parameters, finally investigate the critical behaviors of *significance* in partition transition in detail, and analytically give the formula of the critical points and the phase diagram.

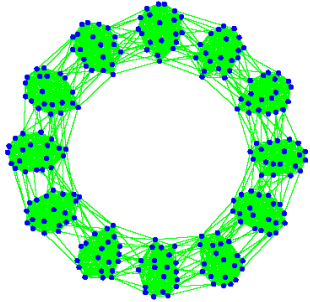

$n = r \cdot n_c$ : Number of vertices in network.

$n_c$ : Number of vertices in each community.

$p_i$ : Probability of linking vertices within community.

$p_o$ : Probability of linking vertices in adjacent communities.

$m = r \cdot n_c^2 p_i + 2r \cdot n_c^2 p_o$ : Number of edges in network.

**Figure 1.** Example of community-loop model networks (drawn by Pajek (http://mrvar.fdv.uni-lj.si /pajek/)) and network parameters.

### 2.1 Definition of model networks

For convenience of theoretical analysis, we constructed a set of community-loop model networks with $r$ communities connected one by one (see Figure 1). For the pre-defined original community partition in the networks, which contains $r$ communities with $n_c$ vertices, the value of *significance* reads,

$$S_{origin} = r \binom{n_c}{2} D(p_1 \| p)$$

$$\approx \frac{r \cdot n_c^2}{2} \left( p_1 \ln \frac{p_1}{p} + (1 - p_1) \ln \frac{1 - p_1}{1 - p} \right) \quad , \quad (2)$$

$$\approx \frac{r \cdot n_c^2}{2} \left( p_i \ln \frac{r \cdot p_i}{p_i + 2 p_o} + (1 - p_i) \ln(1 - p_i) \right)$$

where $p_1 = p_i$, $p = (p_i + 2p_o)/r$, and $1 - p \approx 1$ generally.

In order to analyze the critical behaviors of *significance* in partition transition, we consider a kind of partitions that consists of $r/2$ groups each of which contains 2 adjacent communities and thus has $2n_c$ vertices. Therefore, the value of *significance* for the partition with community merging reads,

$$S_{merge} = \frac{r}{2} \binom{2n_c}{2} D(p_2 \| p)$$

$$\approx r \cdot n_c^2 \left( p_2 \ln \frac{p_2}{p} + (1 - p_2) \ln \frac{1 - p_2}{1 - p} \right) \quad , \quad (3)$$

$$\approx r \cdot n_c^2 \left( \frac{p_i + p_o}{2} \ln \frac{r \cdot \frac{p_i + p_o}{2}}{p_i + 2 p_o} + (1 - \frac{p_i + p_o}{2}) \ln(1 - \frac{p_i + p_o}{2}) \right)$$

where $p_2 = (p_i + p_o)/2$ and $p = (p_i + 2p_o)/r$.

## 2.2 Relationship between significance and network parameters

For the sake of visual illustration, Figures 2 and 3 plot the curves of *significance* with various network parameters, though equations (2) and (3) contain the relations between *significance* and network parameters.

Firstly, suppose $1-p \approx 1$ for large $r$-values, thus $S \propto r \ln r$ for $S_{origin}$ and $S_{merge}$. Figures 2 and 3 also clearly show that the values of *significance*, normalized by $r \cdot n_c^2$, are linearly increase with $r$.

Secondly, for $S_{origin}$, the slops of the curves are affected only by the inner-community link probability $p_i$, while $p_o$ only affects the intercepts of the curves (note the intercepts are also affected by $p_i$). So we can see the family of curves for different $p_i$, which are a series of parallel straight lines for different $p_o$ ( see Figure 2(a)).

Thirdly, for $S_{merge}$, the slops of the curves are affected by $p_i$ and $p_o$, while the intercepts are also. So we see the family of curves for different $p_i$, which contains the straight lines with different slops and intercepts for different $p_o$ ( see Figure 2(b)).

Finally, by comparing the curves for different $p_i$ (see Figure 2), $S_{origin}$ and $S_{merge}$ increase with the increase of $p_i$. By comparing the curves for different $p_o$, $S_{origin}$ decreases with the increase of $p_o$, while $S_{merge}$ increases with the increase of $p_o$. Figure 3 further displays the conclusions more clearly. It also imply that the larger the $p_i$-values, the more difficult the community merging, because the needed $p_o$-values will be larger.

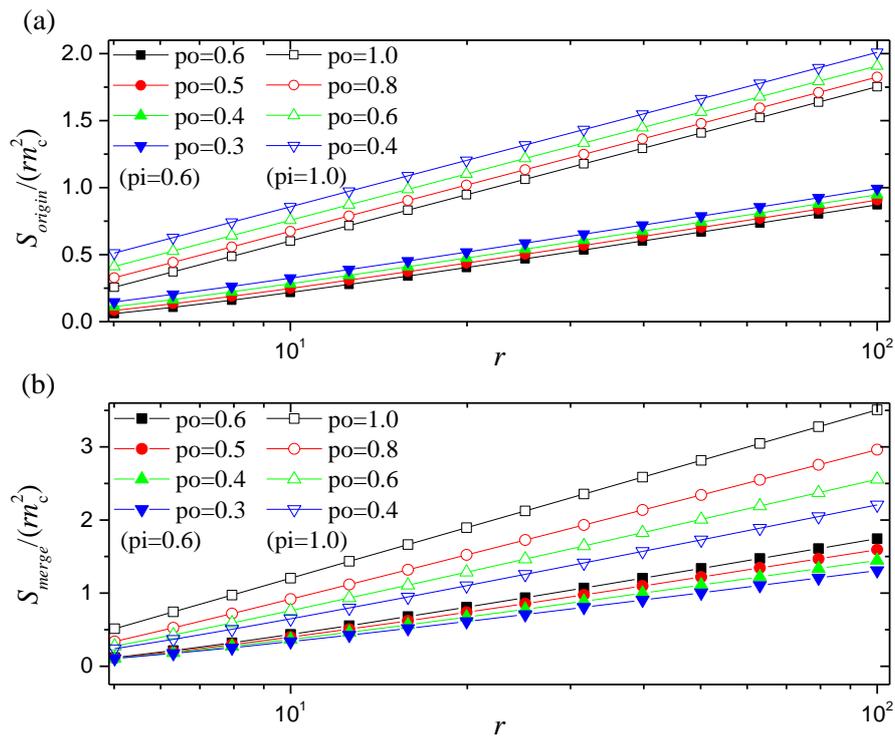

**Figure 2.** Curves of *significance* as a function of *r* for different values of $p_i$ and $p_o$.

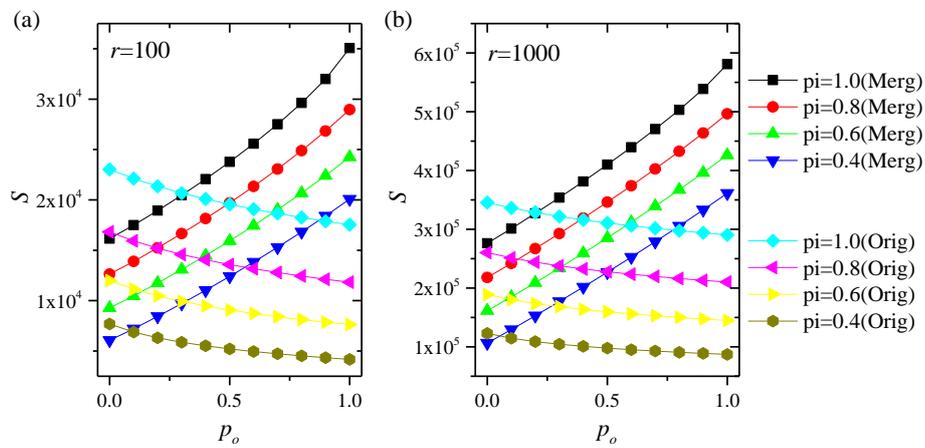

**Figure 3.** For the original partition (Orig) and the partition with community merging (Merg), curves of *significance* as a function of $p_o$ for different $p_i$: (a) *r*=100 and (b) *r*=1000.

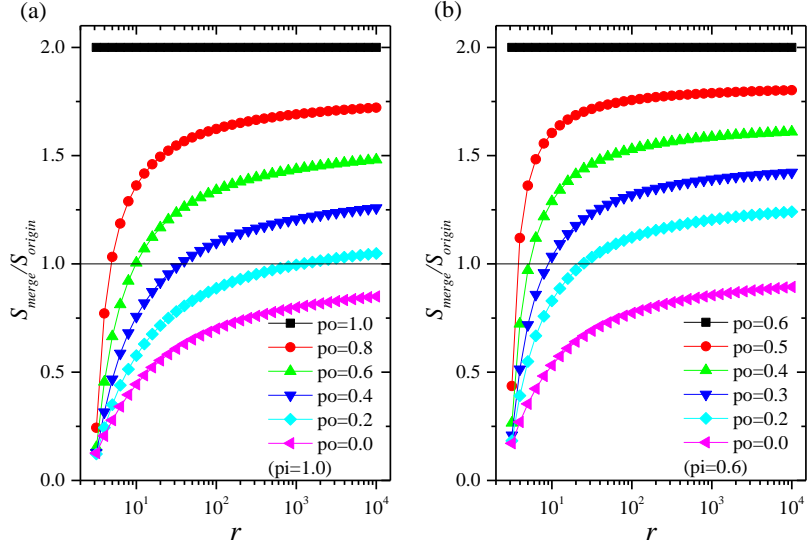

**Figure 4.** Ratio of $S_{merge}$ to $S_{origin}$ as a function of $r$ for different values of $p_i$ and $p_o$.

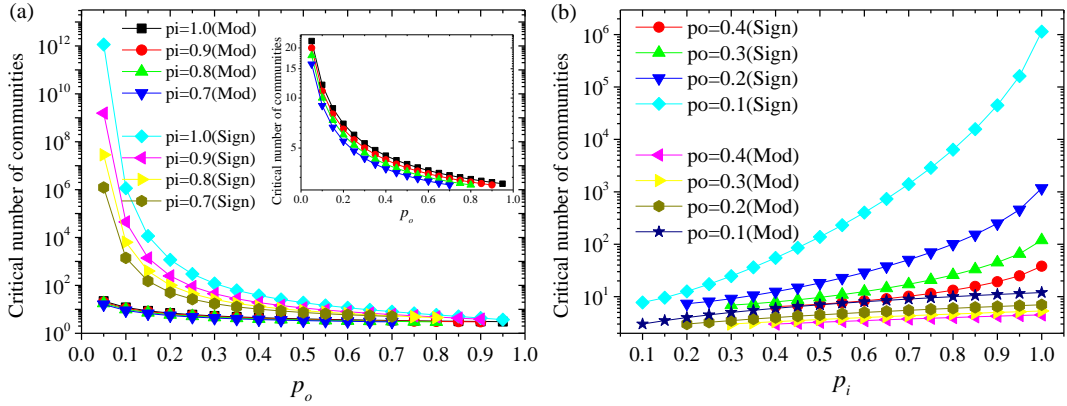

**Figure 5.** The phase diagrams in partition transition. (a) For *significance* (Sign), the critical number of communities as a function of $p_o$, for different $p_i$-values, compared with *modularity* (Mod), and inset graph is more clearly to display the critical number of communities for modularity. (b) The critical number of communities as a function of $p_i$, for different $p_o$-values, compared with modularity.

## 2.3 Critical behaviors in partition transition

In the section, we study the transition from the predefined partition to the partition with community merging. When $S_{merge} - S_{origin} > 0$, the identified partition should change to be the above the partition with community merging while not the pre-defined original partition. Figure 4 shows that $S_{merge}/S_{origin}$ will be greater than 1, when the number $r$ of communities is large enough, and the critical points are different for different $p_o$-values (*e.g.*, from 0.0 to 1.0). As we see that, the smaller the $p_o$-values, the larger the needed $r$-values, meaning that the community merging is more difficult. For smaller $p_i$ (e.g. $p_i = 0.6$), the needed $r$-values decrease correspondingly.

On the basis of the above qualitative analysis, in the following, we give the analytic expression of the critical points in the partition transition. By equations (2) and (3), the critical condition in the transition reads,

$$\left( \binom{2n_c}{2} D(p_2 \parallel p) - 2 \binom{n_c}{2} D(p_1 \parallel p) \right) \geq 0 . \tag{4}$$

**Theorem.** By solving Eq. (4) for $r$, the critical number of communities for *significance* in the partition transition, reads,

$$
\begin{aligned}
r_{critical} &= p' \cdot \exp\left( \frac{2H(p_2) - H(p_1)}{2p_2 - p_1} \right) \\
&= p' \cdot \exp\left( \frac{H(p_1) + 2\Delta H}{p_1 + 2\Delta p} \right) \\
&= p' \cdot \exp\left( \frac{1 + 2\Delta H / H(p_1)}{1 + 2\Delta p / p_1} \cdot \frac{H(p_1)}{p_1} \right)
\end{aligned}
, \tag{5}
$$

where the information entropy $H(x) = -x\ln(x) - (1-x)\ln(1-x)$, $\Delta H = H(p_2) - H(p_1)$, $\Delta p = p_2 - p_1$ and $p' = p_i + 2p_o$. The critical number of communities in the transition is closely related to the changes of the information entropy caused by the inner-link probability in the communities. Moreover, for modularity, $r_{critical} = p_i / p_o + 2$.

**Proof.** Suppose that $n_c - 1 \approx n_c$, $2n_c - 1 \approx 2n_c$, $1-p \approx 1$ for large $r$–values, and define $p = p'/r$. By Eq. (4),

$$0 = \frac{2n_c(2n_c - 1)}{2} D(p_2 \parallel p) - 2 \frac{n_c(n_c - 1)}{2} D(p_1 \parallel p)$$

$$0 = 2\left( p_2 \ln\left( \frac{p_2}{p'} r \right) + (1-p_2)\ln(1-p_2) \right) - \left( p_1 \ln\left( \frac{p_1}{p'} r \right) + (1-p_1)\ln(1-p_1) \right)$$

$$(2p_2 - p_1)\ln(r) = p_1 \ln(p_1) + (1-p_1)\ln(1-p_1) - 2\left( p_2 \ln p_2 + (1-p_2)\ln(1-p_2) \right) + (2p_2 - p_1)\ln p'$$

$$
\begin{aligned}
r &= \exp\left( \frac{1}{2p_2 - p_1} \left( p_1 \ln\left( \frac{p_1}{p'} \right) + (1-p_1)\ln(1-p_1) - 2\left( p_2 \ln\left( \frac{p_2}{p'} \right) + (1-p_2)\ln(1-p_2) \right) \right) + \ln p' \right) \\
&= p' \cdot \exp\left( \frac{2H(P_2) - H(P_1)}{2p_2 - p_1} \right)
\end{aligned}
\tag{6}
$$

For illustration, Figure 5 display the relation between $r_{critical}$ and network parameters. We can see that $r_{critical}$ decreases with the increase of $p_o$. This is reasonable, because the increase of the number of links between communities will make the communities merging more easily. For large $p_o$-values, the $r_{critical}$-values are very small, which are close to the critical values of modularity. However, for small $p_o$-values, the $r_{critical}$-values dramatically increase with the decrease of $p_o$, which is far greater than that of *modularity*. As a result, *significance* generally tends to split the communities in the networks, especially with small inter-community link density, and find more communities than other methods, such as *modularity*. This is confirmed by the experimental results in the next section.

Moreover, we see that for fixing $p_o$-values, the larger the $p_i$-values, the larger the $r_{critical}$-values (see Figure 5(b)). That means that the denser the links inside communities, the more difficult the communities merging. On the whole, the difference between inter- and intra-community link density is easily to result in the disconnecting of communities. The slight link-density inhomogeneity in community is also possible to lead to the split of the community. In some cases, some high link-density regions may be separated from the communities in networks.

### 3. Experimental Results

In this section, we provide a series of comparisons of *significance* with some classical methods such as *Modularity*[26], *Infomap*[12], *Walktrap*[51], *OSLOM*[52] and *LP*[53]) on artificial and real-world networks.

### 3.1 Artificial networks

Firstly, we identify the communities in the above community-loop networks by using Louvain algorithm for significance optimization. Table 1 shows that (1) when the number of pre-defined communities $r$ is large enough, communities merging will appear, e.g. for $p_i = 1.0$ and $p_o = 0.4$; (2) when $p_o$ is large enough, communities merging will appear, e.g. for $p_i = 1.0$ and $r = 128$; (3) the decrease of $p_i$ makes communities more easily merge. Table 2 shows similar results for modularity in the same networks, but *modularity* is more easily to merge the communities in the networks than *significance*. These results are consistent with the above theoretical analysis.

Figure 6 compared the accumulative number of identified communities by different methods in the community-loop networks with different parameters. It confirm that *significance* can identify more communities than other methods. Or say, *significance* has higher resolution in community detection.

However, the high resolution of *significance* may lead to another problem - the excessive splitting of communities. In some cases, it may not be able to identify the community structures, which can be identified by some classical methods. We test a set of examples for this problem. Table 3 shows that the ratio of the number of communities identified by different methods, to the number of predefined communities, in the LFR networks [54]. With the decrease of the mean degree $k_m$ in the networks, the split of communities is getting worse, because of the increase of inhomogeneity inside communities.

**Table 1.** Ratio of the number of identified communities to the number of predefined communities, by *significance*, in the community-loop networks with different network parameters.

| $p_i$ | $p_o$ | Number of pre-defined communities $r$ | | | | | |
|---|---|---|---|---|---|---|---|
| | | 4 | 8 | 16 | 32 | 64 | 128 |
| 1.0 | 0.2 | 1.0 | 1.0 | 1.0 | 1.0 | 1.0 | 1.0 |
| | 0.3 | 1.0 | 1.0 | 1.0 | 1.0 | 1.0 | **0.6** |
| | 0.4 | 1.0 | 1.0 | 1.0 | **0.6** | **0.6** | **0.6** |
| 0.8 | 0.2 | 1.0 | 1.0 | 1.0 | 1.0 | 1.0 | **0.6** |
| | 0.3 | 1.0 | 1.0 | 1.0 | **0.6** | **0.6** | **0.4** |
| | 0.4 | 1.0 | 1.0 | **0.6** | **0.6** | **0.4** | **0.4** |

**Table 2.** Ratio of the number of identified communities to the number of predefined communities, by *modularity*, in the community-loop networks with different network parameters.

| $p_i$ | $p_o$ | Number of pre-defined communities $r$ | | | | | |
|---|---|---|---|---|---|---|---|
| | | 4 | 8 | 16 | 32 | 64 | 128 |
| 1.0 | 0.2 | 1.0 | **0.6** | **0.4** | **0.3** | **0.2** | **0.2** |
| | 0.3 | 1.0 | **0.6** | **0.4** | **0.3** | **0.2** | **0.1** |
| | 0.4 | **0.9** | **0.6** | **0.4** | **0.3** | **0.2** | **0.1** |
| 0.8 | 0.2 | 1.0 | **0.5** | **0.4** | **0.3** | **0.2** | **0.2** |
| | 0.3 | **0.6** | **0.4** | **0.3** | **0.3** | **0.2** | **0.1** |
| | 0.4 | **0.6** | **0.4** | **0.3** | **0.3** | **0.2** | **0.1** |

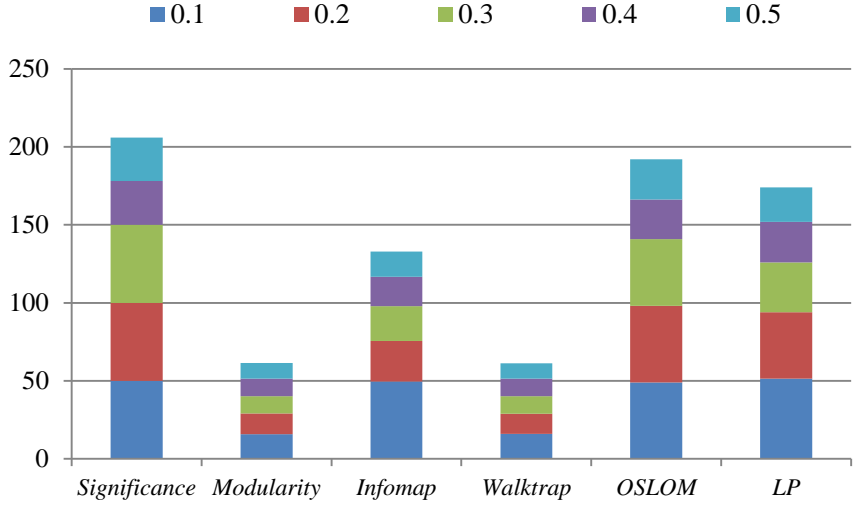

**Figure 6.** The accumulative number of identified communities by different methods (Significance, Modularity, Infomap, Walktrap, OSLOM and LP) in the community-loop networks with $r$=50, $p_i$ = 1 and $p_o$ =0.1, 0.2, 0.3, 0.4 and 0.5.

**Table 3.** Ratio of the number of communities identified by different methods, to the number of predefined communities, in the LFR networks with different values of $k_m$ and $C_{max}$, $N$=500, $k_{max}$=50, $C_{min}$=20, $\mu$=0.1, $\tau_1$=2, and $\tau_1$=2.

| $C_{max}$ | $k_m$ | Significance | Modularity | Infomap | Walktrap | OSLOM | LP |
|-----------|-------|--------------|------------|---------|----------|-------|-----|
|           | 12    | **3.0**      | 1.0        | 1.0     | 1.0      | 1.0   | **2.9** |
| 50        | 16    | **2.0**      | 1.0        | 1.0     | 1.0      | 1.0   | **1.9** |
|           | 20    | **1.4**      | 1.0        | 1.0     | 1.0      | 1.0   | **1.1** |
|           | 12    | **5.6**      | 1.0        | 1.0     | 1.0      | 1.0   | **1.7** |
| 100       | 16    | **3.3**      | 1.0        | 1.0     | 1.0      | 1.0   | **1.3** |
|           | 20    | **2.3**      | 1.0        | 1.0     | 1.0      | 1.0   | **1.0** |

## 3.2 Real-world networks.

Finally, we apply the above methods to a set of real-word networks. In the real-world networks, it is difficult to directly compare the performance of different methods. In Table 4, therefore, we list the number of communities identified by different methods. The results show that *significance* intensively splits the networks into communities. This confirmed that *significance* also tends to generate more communities in the general real-world networks than other methods.

**Table 4.** The number of communities in various real-world networks, identified by different methods.

| Networks | Modularity | Infomap | Walktrap | OSLOM | LP | Significance |
|----------|------------|---------|----------|-------|-----|--------------|
| Dolphin[55] | 5 | 6 | 5 | 2 | 10 | **22** |
| Polbooks [56] | 5 | 5 | 4 | 2 | 6 | **28** |
| Football [57] | 9 | 12 | 10 | 11 | 11 | **15** |
| Jazz [58] | 4 | 6 | 8 | 5 | 3 | **36** |
| C. Elegans neural [59] | 5 | 8 | 22 | 3 | 2 | **67** |
| Email [60] | 11 | 63 | 47 | 24 | 9 | **233** |

**4. Discussion and conclusion**

Community structure extensively exists in various complex networks. Detecting communities (or modules) in complex networks is very important for the research of complex networks. In the past decades, much attention was paid to the development of methods for community detection in complex networks. However, the detailed analysis of the methods' behaviors is also of interest, which could help in understanding the method themselves, and promote the development of more effective methods.

In this paper, we focus on an important measure for estimating the quality of community structures, called *significance*. It was proposed to initially determine significant scale of community structures, but it can also be used as a target function to search the optimal community partitions. We studied the effect of various network parameters on this measure in detail, analyzed the critical behaviors of it in partition transition, and analytically gave the formula of the critical points and the phase diagram. The results were confirmed on artificial and real-world networks, and a series of comparisons with some classical methods were also given.

The difference between inter- and intra-community link density is crucial to the disconnecting or splitting of communities in networks. The results shown that the critical number $r_{critical}$ of communities in partition transition is to increase dramatically with the decrease of the inter-community link density for each intra-community link density. When the inter-community link density is very large, the $r_{critical}$-value is very small, which is close to but still large than that of *modularity*, but when the inter-community link density becomes small, the $r_{critical}$-value quickly increases, and is far greater than that of *modularity*.

On the whole, it was shown that *significance* tends to split the communities in the networks, and find more communities than other methods, such as *modularity*. So it generally has higher resolution in community detection than many other methods, but it also may lead to the problem of excessive splitting of communities. In some cases, the low link-density inhomogeneity in community is also possible to lead to the split of the community. It is still open issue how to find the appropriate balance between the high resolution and excessive splitting in community detection.

Finally, we expect that the above detailed analysis could be helpful for the understanding of the behaviors of *significance* in community detection and provide useful insight into designing effective methods for detecting communities in complex networks.

**Acknowledgement**

This work was supported by the construct program of the key discipline in Hunan province, the Scientific Research Project of Hunan Provincial Health and Family Planning Commission of China (Grant No. C2017013), the Scientific Research Fund of Education Department of Hunan Province (Grant Nos. 14C0112 and 14B024), the Department of Education of Hunan Province (Grant No. 15A023), the Hunan Provincial Natural Science Foundation of China (Grant No. 2015JJ6010), the Hunan Provincial Natural Science Foundation of China (Grant No. 13JJ4045), and the National Natural Science Foundation of China (Grant No. 11404178 and Grant No. 71401194).